\documentstyle[preprint,tighten,aps]{revtex}
\begin{document} 
\draft
\preprint{IASSNS-HEP-96/24,PUPT-1602}
\date{March 1996}
\title{General Rotating Five Dimensional Black Holes of Toroidally
Compactified Heterotic String}
\author{Mirjam Cveti\v c$^1$
\thanks{On sabbatic leave from the University of Pennsylvania.
E-mail address: cvetic@sns.ias.edu}
and Donam Youm$^2$
\thanks{On leave from the University of Pennsylvania. 
E-mail addresses: youm@pupgg.princeton.edu; youm@sns.ias.edu}}
\address{$^1$ School of Natural Science, Institute for Advanced
Study\\ Olden Lane, Princeton, NJ 08540 \\ and \\
$^2$ Physics Department, Joseph Henry Laboratories\\ 
Princeton University, Princeton, NJ 08544}
\maketitle
\begin{abstract}
{We present the most general rotating black hole solution of
five-dimensional $N=4$ superstring vacua that conforms to the 
``no hair theorem''.  It is chosen to be parameterized in terms of  
massless fields of the toroidally compactified heterotic string.
The solutions are obtained by performing a subset of $O(8,24)$ 
transformations, {\it i.e.}, symmetry transformations of the 
effective three-dimensional action for stationary solutions, 
on the five-dimensional (neutral) rotating solution parameterized 
by the mass $m$ and two rotational parameters $l_1$ and $l_2$.  
The explicit form of the generating solution is determined by 
three $SO(1,1)\subset O(8,24)$ boosts, which specify two electric 
charges $Q_1^{(1)},\ Q_{2}^{(2)}$ of the Kaluza-Klein and two-form 
$U(1)$ gauge fields associated with the same compactified direction, and 
the charge $Q$ (electric charge of the vector field, whose field 
strength is dual to the field strength of the five-dimensional 
two-form field).  The general solution, parameterized by 27 charges, 
two rotational parameters and the ADM mass compatible with the 
Bogomol'nyi bound, is obtained by imposing 
$[SO(5)\times SO(21)]/[SO(4)\times SO(20)]\subset O(5,21)$
transformations, which do not affect the five-dimensional space-time.  
We also analyze the deviations from the BPS-saturated limit.}
\end{abstract}

\section{Introduction}

Recently, certain black holes in string theory have attracted
an overwhelming attention due to the fact that it has now become 
feasible to address their thermal properties in terms of
microscopic degrees of freedom associated with the $D$-brane
configuration
\footnote{For a review on $D$-brane physics see Ref. \cite{CJP}.}, 
describing a particular black hole.  
In particular, the microscopic entropy of certain five-dimensional  
BPS-saturated static \cite{SV,CM} and rotating \cite{BMPV} 
black holes as well as that for specific infinitesimal  
deviations from the BPS-saturated limit for static \cite{HS} 
and rotating \cite{BLMPSV} six-dimensional black strings 
(or five-dimensional black holes upon dimensionally reducing 
on $S^1$) have been provided
\footnote{Recently, the microscopic entropy in terms of the 
degrees of freedom of the corresponding $D$-brane configurations 
for certain (four-parameter) four-dimensional BPS-saturated 
black holes has been given \cite{MSTR,JKM}, while an earlier  
complementary approach was initiated in Ref. \cite{LW} 
with further elaboration in Refs.\cite{CT,T}.}.

Interestingly, the ``$D$-brane technology'' allows one to calculate 
the microscopic entropy of certain types of black holes whose 
explicit classical configurations have  either not been constructed, 
yet,  or have been constructed only for a specific assignment of charges.  
The purpose of this paper is to fill such a void for the 5-dimensional 
rotating black hole solutions.  Namely, we shall present an explicit form 
of the generating solution for the general rotating black hole solution 
of five-dimensional $N=4$ superstring vacua, that conforms to the 
``no hair theorem''
\footnote{The content of the conjecture no-hair theorem states that 
the ADM mass, the angular momenta and a set of conserved (electric 
and magnetic) charges uniquely specifies classical black hole solution.  
In the case under consideration in this paper, the black hole solutions 
are uniquely parameterized by the ADM mass (which can be traded for the 
non-extremality parameter $m>0$), two angular momenta and 27 (conserved) 
electric charges.}
\cite{HAIR}.   We choose to parameterize the general 
solution in terms of massless fields of the heterotic string 
compactified on a five-torus ($T^5$)
\footnote{Equivalent parameterization is possible 
(due to string-string duality) in terms of fields of Type IIA 
compactified on $K3\times S^1$ or $T$-dualized Type IIB 
string.  In the latter cases the generating solution has a map 
onto Ramond-Ramond (RR) charges and thus an interpretation in 
terms of $D$-brane configurations.}.  
 
We use a generating technique by performing a subset of $O(8,24)$ 
transformations, {\it i.e.}, symmetry transformations of the 
effective three-dimensional action for stationary solutions, 
on the five-dimensional (neutral) rotating solution parameterized 
by the mass $m$ and two rotating parameters $l_1$ and
$l_2$.  The explicit form of the generating solution is determined 
by three $SO(1,1)\subset O(8,24)$ boosts $\delta_{e1},\ 
\delta_{e2}$, and $\delta_e$, which specify respectively the two 
electric charges $Q_1^{(1)},\ Q_1^{(2)}$ of the two $U(1)$ gauge 
fields, {\it i.e.}, the Kaluza-Klein $A_{\mu 1}^{(1)}$ and the 
two-form $A_{\mu 1}^{(2)}$ gauge fields associated with the 
first compactified direction, and the charge $Q$, {\it i.e.}, 
the electric charge of the vector field, whose field strength 
is dual to the field strength $H_{\mu\nu\rho}$ 
of the five-dimensional two form field $B_{\mu\nu}$.  
The general solution, parameterized by 27 charges, two 
rotational parameters and the ADM mass compatible with the 
Bogomol'nyi bound, is obtained by imposing on the generating 
solution $[SO(5)\times SO(21)]/[SO(4)\times SO(20)]\subset O(5,21)$ 
transformations, which do not affect the five-dimensional space-time.  
Here the $O(5,21)$ symmetry corresponds to the $T$-duality 
symmetry of the five-dimensional toroidally compactified 
heterotic string vacuum.

Our results reproduce the generating solution for the 
BPS-saturated  static \cite{SV,T,CM} and rotating\cite{BMPV,T} 
black holes (as well as specific deviations \cite{HS,BLMPSV} from 
the BPS limit) as special examples.  The mass formula and the area 
of the horizon for the general BPS-saturated black hole at generic 
points of the $N=4$ moduli and coupling space is written 
in terms of one rotational  parameter, 27 (conserved) quantized 
charges as well as the arbitrary asymptotic values of 115 moduli 
fields and the dilaton.   Again, the area of the horizon is  
independent of the moduli and the dilaton couplings. 

The paper is organized as follows.  In Section II we write down the 
explicit form of the toroidally compactified effective heterotic 
string action in $D$-dimensions, with the emphasis on five-dimensional 
action and three-dimensional effective action suitable for 
describing the stationary solutions in higher dimensions.  
In Section III we give the explicit form of the generating solution, 
specified by six parameters (mass, three charges and two rotational 
parameters) and give a subset of $T$-duality transformations which 
allow one to write the general solution in terms of the mass, 
27 (quantized) charges and two rotational parameters, as well as 
arbitrary asymptotic values of moduli and dilaton couplings. 
In Section IV we concentrate on the BPS-saturated solution, 
and properties of the deviations from the BPS limit.  In particular 
we address the mass formula and the area of the horizon.  

\section{Action of Heterotic String on Tori} 

In this chapter, we shall discuss effective actions of the 
massless bosonic fields of heterotic string compactified on 
tori.  The starting point is the following zero-slope limit 
effective action of the heterotic string with zero mass fields 
given by the graviton $\hat{G}_{MN}$, the two form field 
$\hat{B}_{MN}$, $U(1)^{16}$ part $\hat{A}^I_M$ of the 
ten-dimensional gauge group, and the dilaton $\hat{\Phi}$ 
($M,N=0,1,...,9$, $I=1,...,16$)
\footnote{We follow the convention spelled out in Refs. 
\cite{MSCH,SENFOUR}.}:  
\begin{equation}
{\cal L} = {1\over {16\pi G_{10}}}\sqrt{-\hat{G}}\,e^{-\hat{\Phi}}
[{\cal R}_{\hat{G}} + \hat{G}^{MN}\partial_M \hat{\Phi} \partial_N 
\hat{\Phi} - {\textstyle {1\over {12}}}\hat{H}_{MNP}\hat{H}^{MNP}-
{\textstyle{1\over 4}}\hat{F}^I_{MN}\hat{F}^{I\,MN}],
\label{10daction}
\end{equation}
where $G_{10}$ is the ten-dimensional Newton's constant, which in this 
paper is chosen to be $8\pi^6$, $\hat G \equiv {\rm det}\,\hat{G}_{MN}$, 
${\cal R}_{\hat G}$ is the Ricci scalar of the metric $\hat{G}_{MN}$, 
$\hat{F}^I_{MN} = \partial_M \hat{A}^I_N - \partial_N \hat{A}^I_M$ and 
$\hat{H}_{MNP} = \partial_M \hat{B}_{NP}-{1\over 2}\hat{A}^I_M 
\hat{F}^I_{NP} + {\rm cyc.\ perms.}$ are the field strengths of 
$\hat{A}^I_M$ and $\hat{B}_{MN}$, respectively.  We choose the mostly 
positive signature convention $(-++\cdots +)$ for the metric $\hat{G}_{MN}$. 

\subsection{Effective Action of Heterotic String Compactified  
on a $(10-D)$-Torus}

The compactification of the extra $(10-D)$ spatial coordinates 
on a $(10-D)$-torus can be achieved by choosing the following 
Abelian Kaluza-Klein Ansatz for the ten-dimensional metric
\begin{equation}
\hat{G}_{MN}=\left(\matrix{e^{a\varphi}g_{{\mu}{\nu}}+
G_{{m}{n}}A^{(1)\,m}_{{\mu}}A^{(1)\,n}_{{\nu}} & A^{(1)\,m}_{{\mu}}
G_{{m}{n}}  \cr  A^{(1)\,n}_{{\nu}}G_{{m}{n}} & G_{{m}{n}}}\right),
\label{4dkk}
\end{equation}
where $A^{(1)\,m}_{\mu}$ ($\mu = 0,1,...,D-1$; 
$m=1,...,10-D$) are $D$-dimensional Kaluza-Klein $U(1)$ gauge fields,  
$\varphi \equiv \hat{\Phi} - {1\over 2}{\rm ln}\,{\rm det}\, G_{mn}$ 
is the $D$-dimensional dilaton field, and $a\equiv {2\over{D-2}}$.   
Then, the affective action is specified by the following massless 
bosonic fields: the (Einstein-frame) graviton $g_{\mu\nu}$, the 
dilaton $\varphi$, $(36-2D)$ $U(1)$ gauge fields ${\cal A}^i_{\mu}
\equiv (A^{(1)\,m}_{\mu},A^{(2)}_{\mu\,m},A^{(3)\,I}_{\mu})$ defined 
as $A^{(2)}_{\mu\,m} \equiv \hat{B}_{\mu m}+\hat{B}_{mn}
A^{(1)\,n}_{\mu}+{1\over 2}\hat{A}^I_mA^{(3)\,I}_{\mu}$, 
$A^{(3)\,I}_{\mu} \equiv \hat{A}^I_{\mu} - \hat{A}^I_m 
A^{(1)\,m}_{\mu}$, the two-form field $B_{\mu\nu} \equiv 
\hat{B}_{\mu\nu}-\hat{B}_{mn}A^{(1)\,m}_{\mu}A^{(1)\,n}_{\nu} 
-{1\over 2}(A^{(1)\,m}_{\mu}A^{(2)}_{\nu\,m}-A^{(1)\,m}_{\nu}
A^{(2)}_{\mu\,m})$, and the following symmetric $O(10-D,26-D)$ 
matrix of the scalar fields (moduli):
\begin{equation}
M=\left ( \matrix{G^{-1} & -G^{-1}C & -G^{-1}a^T \cr 
-C^T G^{-1} & G + C^T G^{-1}C +a^T a & C^T G^{-1} a^T 
+ a^T \cr -aG^{-1} & aG^{-1}C + a & I + aG^{-1}a^T} 
\right ), 
\label{modulthree}
\end{equation} 
where $G \equiv [\hat{G}_{mn}]$, $C \equiv [{1\over 2}
\hat{A}^{(I)}_{{m}}\hat{A}^{(I)}_{n}+\hat{B}_{mn}]$ and 
$a \equiv [\hat{A}^I_{{m}}]$ are defined in terms of the 
internal parts of ten-dimensional fields.  Then the effective 
$D$-dimensional effective action takes the form:
\begin{eqnarray}
{\cal L}&=&{1\over{16\pi G_D}}\sqrt{-g}[{\cal R}_g-{1\over (D-2)}
g^{\mu\nu}\partial_{\mu}\varphi\partial_{\nu}\varphi+{1\over 8}
g^{\mu\nu}{\rm Tr}(\partial_{\mu}ML\partial_{\nu}ML)-{1\over{12}}
e^{-2a\varphi}g^{\mu\mu^{\prime}}g^{\nu\nu^{\prime}}
g^{\rho\rho^{\prime}}H_{\mu\nu\rho}H_{\mu^{\prime}\nu^{\prime}
\rho^{\prime}} \cr
&-&{1\over 4}e^{-a\varphi}g^{\mu\mu^{\prime}}g^{\nu\nu^{\prime}}
{\cal F}^{i}_{\mu\nu}(LML)_{ij}
{\cal F}^{j}_{\mu^{\prime}\nu^{\prime}}],
\label{effaction}
\end{eqnarray}
where $G_D$ is the $D$-dimensional Newton's constant
\footnote{The Newton's constant $G_D$ in $D$-dimensions can be expressed 
in terms of the 10-dimensional one $G_{10}$ as $G_{10}=(2\pi
\sqrt{\alpha^{\prime}})^{10-D}G_D$, where in this paper we choose 
$\alpha^{\prime}=1$.},  
$g\equiv {\rm det}\,g_{\mu\nu}$, ${\cal R}_g$ is the Ricci 
scalar of $g_{\mu\nu}$, ${\cal F}^i_{\mu\nu} = \partial_{\mu} 
{\cal A}^i_{\nu}-\partial_{\nu} {\cal A}^i_{\mu}$ are the 
$U(1)^{36-2D}$ gauge field strengths, and $H_{\mu\nu\rho} \equiv 
(\partial_{\mu}B_{\nu\rho}-{1\over 2}{\cal A}^i_{\mu}L_{ij}
{\cal F}^j_{\nu\rho}) + {\rm cyc.\ perms.\ of} \mu , \nu , \rho$
is the field strength of the two-form field $B_{\mu\nu}$.  
Here, in the above we have chosen a metric $L$ of $O(10-D,26-D)$ 
group to be:
\begin{equation}
L =\left ( \matrix{0 & I_{10-D}& 0\cr I_{10-D} & 0& 0 \cr 0 & 0 &  
I_{26-D}} \right ) ,
\label{invmet}
\end{equation}
where $I_n$ denotes the $n\times n$ identity matrix.
The $D$-dimensional effective action (\ref{effaction}) is 
invariant under the $O(10-D,26-D)$ transformations ($T$-duality) 
\cite{MSCH,SENFOUR}:
\begin{equation}
M \to \Omega M \Omega^T ,\ \ \ {\cal A}^i_{\mu} \to \Omega_{ij}
{\cal A}^j_{\mu}, \ \ \ g_{\mu\nu} \to g_{\mu\nu}, \ \ \ 
\varphi \to \varphi, \ \ \ B_{\mu\nu} \to B_{\mu\nu}, 
\label{tdual}
\end{equation}
where $\Omega$ is an $O(10-D,26-D)$ invariant matrix, {\it i.e.}, 
$\Omega^T L \Omega = L$.  

In particular for $D=5$ the effective action is specified by the 
graviton, 116 scalar fields (115 moduli fields in the matrix $M$ and 
the dilaton $\varphi$), 26 $U(1)$ gauge fields, and the 
field strength $H_{\mu\nu\rho}$ of the two form field $B_{\mu\nu}$.  
By the duality transformation:
 \begin{equation}
H^{\mu\nu\rho}=-{e^{4\varphi/3}\over{2!\sqrt{-g}}}
\varepsilon^{\mu\nu\rho\lambda\sigma}F_{\lambda\sigma}, 
\end{equation}
$H_{\mu\nu\rho}$ can be related to the field strength 
$F_{\mu\nu}$ of the gauge field $A_\mu$, and therefore in 
five-dimensions black hole solution carries extra $U(1)$ charge 
associated with the dual of the three-form field strength. 

The $T$-duality symmetry of the effective action is $O(5,21)$.  
The equations of motion and Bianchi identities are also 
invariant under the $SO(1,1)$ symmetry
\footnote{For a review of the supergravity duality groups in 
different dimensions see, {\it e.g.}, Ref. \cite{SS}.}:
\begin{equation}
\varphi \to  \varphi+\beta, \ \ H_{\mu\nu\rho} \to {\rm 
e}^{-2\beta/3}H_{\mu\nu\rho}, \ \  
{\cal F}_{\mu\nu}^i\to  {\rm e}^{+\beta/3}{\cal F}_{\mu\nu}^i, \ \ 
M\to M, \ \ g_{\mu\nu}\to g_{\mu\nu}, 
\label{sduality}
\end{equation}
where $\beta$ is an $SO(1,1)$ boost parameter.  Such an $SO(1,1)$ 
symmetry can be used to set the asymptotic value of the 
three-dimensional dilaton to be $\varphi_{\infty}=0$. 

At the quantum level, the parameters of both $O(5,21)$ and $SO(1,1)$ 
symmetry transformations become integer-valued.
  
\subsection{Three-Dimensional Effective Action}

The generating five-dimensional {\it charged} rotating solution
is obtained by imposing $SO(1,1)\subset O(8,24)$ boost transformations 
on the five-dimensional neutral solution.  Here $O(8,24)$ is the 
symmetry of the three-dimensional effective action of the heterotic 
string theory with Killing coordinates (which we choose to be the 
time coordinate $t$, one of the angles of the rotations (which we 
choose to be $\psi$), and the internal coordinates of the 
five-dimensional space-time compactified on a seven-torus.  
The $O(8,24)$ contains as subsets three-dimensional $O(7,23)$ 
$T$-duality symmetry discussed in the previous section and 
four-dimensional $SL(2)$ $S$-duality symmetry. 
Such an enhancement of the duality symmetry is possible due to the 
fact that in three-dimensions the field strength of a one-form 
field is dual to a scalar as discussed below. 
Such an action is given by: 
\begin{equation}
{\cal L} = {\textstyle {1\over 4}}\sqrt{-h}\,[{\cal R}_h + 
{\textstyle {1\over 8}}h^{\bar{\mu}\bar{\nu}}{\rm Tr}
(\partial_{\bar{\mu}}{\cal M}{\bf L}
\partial_{\bar{\nu}}{\cal M}{\bf L})],
\label{3daction}
\end{equation}
where $h\equiv {\rm det}\,h_{\bar{\mu}\bar{\nu}}$ and ${\cal R}_h$ 
is the Ricci scalar of the three-dimensional metric 
$h_{\bar{\mu}\bar{\nu}}$ ($\bar{\mu},\bar{\nu} = r,\theta,\phi$).  
Here, we use ``bars'' to denote fields in three-dimensions.  
${\cal M}$ is a symmetric $O(8,24)$ matrix of three-dimensional 
scalar fields defined as 
\begin{equation}
{\cal M} = \left ( \matrix{\bar{M}-e^{2\bar\varphi}\psi\psi^T &
e^{2\bar\varphi}\psi & \bar{M}\bar{L}\psi-{1\over 2}
e^{2\bar\varphi}\psi(\psi^T \bar{L} \psi) 
\cr 
e^{2\bar\varphi}\psi^T & -e^{2\bar\varphi} & {1\over 
2}e^{2\bar\varphi}\psi^T \bar{L}\psi \cr 
\psi^T \bar{L}\bar{M} - {1\over 2}e^{2\bar\varphi}\psi^T
(\psi^T \bar{L}\psi) & 
{1\over 2}e^{2\bar\varphi}\psi^T \bar{L}\psi & -e^{-2\bar\varphi} 
+ \psi^T \bar{L}\bar{M}\bar{L}\psi - 
{1\over 4}e^{2\bar\varphi}(\psi^T \bar{L}\psi)^2}
 \right ), 
\label{modultwo}
\end{equation}
where $\bar{M}$ is the $O(7,23)$ symmetric moduli field and 
$\bar\varphi$ is the three-dimensional dilaton.  Here, 
$\psi \equiv [\psi^i]$ is the scalars dual to the 
three-dimensional $U(1)$ gauge fields $\bar{\cal A}^i_{\mu}$ 
\cite{SENTHREE}:
\begin{equation}
\sqrt{-h}e^{-2{\bar\varphi}}h^{\bar{\mu}\bar{\mu}^{\prime}}
h^{\bar{\nu}\bar{\nu}^{\prime}}(\bar{M}\bar{L})_{\bar{i}\bar{j}}
\bar{{\cal F}}^{\bar{j}}_{\bar{\mu}^{\prime}\bar{\nu}^{\prime}} = 
\epsilon^{\bar{\mu}\bar{\nu}\bar{\rho}}
\partial_{\bar{\rho}}\psi^{\bar{i}}.
\label{dual}
\end{equation}

The action is manifestly invariant under the $O(8,24)$ 
transformations:
\begin{equation}
{\cal M} \to {\bf \Omega} {\cal M} {\bf \Omega}^T, \ \ \ \ 
h_{\bar{\mu}\bar{\nu}} \to h_{\bar{\mu}\bar{\nu}}, 
\label{o824}
\end{equation}
where ${\bf \Omega} \in O(8,24)$, {\it i.e.},
\begin{equation}
{\bf \Omega}{\bf L}{\bf\Omega}^T={\bf L } , \ \  \
 {\bf L} = \left (\matrix{\bar{L} & 0 & 0 \cr 0 & 0 & 1 
\cr 0 & 1 & 0}\right ).
\label{3dbfo}
\end{equation}

\section{General Solution}

In this Section we briefly spell out the solution generating 
technique
\footnote{A general approach of generating charged solutions
from the charge neutral solutions within the toroidally compactified 
heterotic string is developed in Ref.\cite{SENBH}.}, 
and the explicit sequence of symmetry transformations 
applied on the five-dimensional neutral rotating solution.  
Then, we give the explicit form of the generating solution.  
In the last subsection we give explicit subset of
five-dimensional $T$-duality ($O(5,21)$) transformations. 

\subsection{Solution Generating Technique}

An arbitrary asymptotic value of the scalar field matrix ${\cal M}$ 
can be transformed into the form 
\begin{equation}
{\cal M}_{\infty} = {\rm diag}(-1,\overbrace{1,...,1}^6,-1,
\overbrace{1,...,1}^{22},-1,-1) \equiv I_{4,28}
\label{asympt}
\end{equation}
by imposing an $O(8,24)$ transformation, {\it i.e.}, 
${\cal M}_{\infty} \to {\bf \Omega} {\cal M}_{\infty}{\bf 
\Omega}^T = I_{4,28}$ (${\bf \Omega} \in O(8,24)$).  Thus, 
without loss of generality we can confine the analysis by 
choosing the asymptotic value of the form to be ${\cal M}_\infty
=I_{4,28}$ and, then, obtain the solutions with arbitrary 
values of ${\cal M}_\infty$ by undoing the above constant 
$O(8,24)$ transformation
\footnote{In the case of the five-dimensional configurations with 
the asymptotically flat five-dimensional space-time metric, the 
subset of such  $O(8,24)$ transformations that preserves the 
asymptotic flatness of the five-dimensional space-time corresponds 
to the constant $O(5,21)\times SO(1,1)$ transformations 
(see Eqs. (\ref{tdual}) and (\ref{sduality})), which allow one 
to bring the asymptotic values of the moduli matrix $M_\infty =
I_{26}$ and the dilaton field $\varphi_\infty=0$ into arbitrary 
asymptotic values.  At the same time the physical charges are 
written in terms of {\it quantized} charges, which are ``dressed'' 
by the asymptotic values of the moduli and dilaton.}.
Then, the subset of $O(8,24)$ transformations that preserves the 
asymptotic value ${\cal M}_\infty =I_{4,28}$ is $SO(8) \times 
SO(24)$.

The starting point for obtaining the generating five-dimensional 
charged rotating solution is the five-dimensional neutral 
solution with two rotational parameters $l_{1,2}$ \cite{MP}.  
We choose to write it in the following form:
\begin{eqnarray}
ds^2 &=& -{{r^2+l^2_1{\rm cos}^2 \theta + l^2_2{\rm sin}^2 \theta -2m}
\over {r^2+l^2_1{\rm cos}^2 \theta + l^2_2{\rm sin}^2 \theta}}dt^2 +
{{r^2(r^2+l^2_1{\rm cos}^2 \theta + l^2_2{\rm sin}^2 \theta)}\over
{(r^2+l^2_1)(r^2+l^2_2)-2mr^2}}dr^2 \cr 
&+& (r^2+l^2_1{\rm cos}^2 \theta + 
l^2_2{\rm sin}^2 \theta)d\theta^2 
+ {{4ml_1l_2{\rm sin}^2 \theta {\rm cos}^2\theta} \over
{r^2+l^2_1{\rm cos}^2 \theta + l^2_2{\rm sin}^2 \theta}}d\phi d\psi \cr
&+&{{{\rm sin}^2 \theta}\over {r^2+l^2_1{\rm cos}^2 \theta + 
l^2_2{\rm sin}^2 \theta}}[(r^2+l^2_1)(r^2+l^2_1{\rm cos}^2 \theta + 
l^2_2{\rm sin}^2 \theta)+2ml^2_1{\rm sin}^2\theta]d\phi^2 \cr
&+&{{{\rm cos}^2 \theta}\over {r^2+l^2_1{\rm cos}^2 \theta + 
l^2_2{\rm sin}^2 \theta}}[(r^2+l^2_2)(r^2+l^2_1{\rm cos}^2 \theta + 
l^2_2{\rm sin}^2 \theta)+2ml^2_2{\rm cos}^2\theta]d\psi^2 \cr
&-&{{4ml_1{\rm sin}^2\theta}\over{r^2+l^2_1{\rm cos}^2 \theta + 
l^2_2{\rm sin}^2 \theta}}dtd\phi 
-{{4ml_2{\rm cos}^2\theta}\over{r^2+l^2_1{\rm cos}^2 \theta + 
l^2_2{\rm sin}^2 \theta}}dtd\psi.
\label{5dkerr}
\end{eqnarray}
To apply $SO(1,1) \subset O(8,22)$ boosts on the Kerr solution 
(\ref{5dkerr}), one has to rewrite (\ref{5dkerr}) in terms of the 
three-dimensional fields in (\ref{3daction}) by performing the 
standard Kaluza-Klein dimensional reduction, using the the 
corresponding five-dimensional Ansatz of (\ref{4dkk}), and, then, 
by performing the duality transformation (\ref{dual}):
\begin{eqnarray}
h_{\bar{\mu}\bar{\nu}}dx^{\bar{\mu}\bar{\nu}}&=&
{{[(r^2+l^2_2)(r^2+l^2_1{\rm cos}^2\theta +l^2_2{\rm sin}^2
\theta -2m)+2ml^2_2{\rm cos}^2\theta]r^2{\rm cos}^2\theta}\over
{r^2(r^2+l^2_1+l^2_2-2m)+l^2_1l^2_2}}dr^2 \cr 
&+&[(r^2+l^2_2)(r^2+l^2_1{\rm cos}^2\theta +l^2_2{\rm sin}^2
\theta -2m)+2ml^2_2{\rm cos}^2\theta]{\rm cos}^2\theta d\theta^2 \cr
&+&[r^2(r^2+l^2_1+l^2_2-2m)+l^2_1l^2_2]{\rm cos}^2\theta 
{\rm sin}^2\theta d\phi^2, 
\cr
\psi_8&=&{{2ml_1{\rm cos}^2\theta}\over 
{r^2+l^2_1{\rm cos}^2\theta +l^2_2{\rm sin}^2\theta}}, \ \ \ 
\psi_9=-{{2ml_1l_2{\rm cos}^4\theta}\over 
{r^2+l^2_1{\rm cos}^2\theta +l^2_2{\rm sin}^2\theta}},\cr 
\bar{g}_{11}&=&-{{r^2+l^2_1{\rm cos}^2\theta +l^2_2
{\rm sin}^2\theta-2m}
\over {r^2+l^2_1{\rm cos}^2\theta+l^2_2{\rm sin}^2\theta}}, \ \ 
\bar{g}_{12}=-{{2ml_2{\rm cos}^2\theta} \over
{r^2+l^2_1{\rm cos}^2\theta+l^2_2{\rm sin}^2\theta}},\cr
\bar{g}_{22}&=&{{{\rm cos}^2\theta[(r^2+l^2_2)(r^2+l^2_1
{\rm cos}^2\theta+l^2_2{\rm sin}^2\theta)+2ml^2_2{\rm cos}^2\theta]} 
\over {r^2+l^2_1{\rm cos}^2\theta+l^2_2{\rm sin}^2\theta}},\cr
e^{-2\bar\varphi}&=&{{(r^2+l^2_2)(r^2+l^2_1{\rm cos}^2\theta +
l^2_2{\rm sin}^2\theta -2m)+2ml^2_2{\rm cos}^2\theta}
\over {r^2+l^2_1{\rm cos}^2\theta+l^2_2{\rm sin}^2\theta}}
{\rm cos}^2\theta.
\label{3dkerr}
\end{eqnarray}

We shall proceed with the following sequence of the symmetry 
transformations on (\ref{3dkerr}).  In order to obtain the explicit 
form of the generating charged rotating solution, we shall apply 
three necessary $SO(1,1)\subset O(8,24)$ boost transformations on 
the neutral rotating solution (\ref{3dkerr}), which would yield the 
generating solution specified by the mass, {\it three} charge 
parameters, and two rotational parameters.  We then apply on 
the generating solution the $[SO(5)\times SO(21)]/[SO(4)\times 
SO(20)]$ transformations, {\it i.e.}, the subsets of $T$-duality 
transformations, which do not affect the five-dimensional 
space-time (and preserve the asymptotic values $M_\infty=I_{26}$ 
and $\varphi_\infty=0$), however, they introduce 4+20 additional 
charge parameters.  Thus, one in turn obtains the most general 
configuration specified by the mass, 27 electric charges and two 
rotational parameters.

\subsection{Generating Solution}

The generating five-dimensional charged rotating solution can 
be obtained by applying the following three $SO(1,1) \subset 
O(8,24)$ boosts on the three-dimensional scalar matrix $\cal M$, 
specified by the neutral rotating  solution (\ref{3dkerr}).  
The boost transformation ${\bf \Omega}_{e}$ has the following 
form:
\begin{equation}
{\bf \Omega}_{e}\equiv\left (\matrix{1&\cdot&\cdot&\cdot&\cdot&\cdot&
\cdot\cr
\cdot& {\rm cosh}\delta_{e}&\cdot&\cdot&\cdot&-{\rm sinh}
\delta_{e}&\cdot\cr 
\cdot&\cdot&I_6&\cdot&\cdot&\cdot&\cdot \cr
\cdot&\cdot&\cdot&{\rm cosh}\delta_{e}&\cdot&\cdot&{\rm sinh}
\delta_{e}\cr
\cdot&\cdot&\cdot&\cdot&I_{21}&\cdot&\cdot\cr
\cdot&-{\rm sinh}\delta_{e}&\cdot&\cdot&\cdot&{\rm cosh}
\delta_{e}&\cdot\cr
\cdot&\cdot&\cdot&{\rm sinh}\delta_{e}&\cdot&\cdot&{\rm cosh}
\delta_{e}}\right ), 
\label{boost}
\end{equation}
where the dots denote the corresponding zero entries and 
${\bf \Omega}_{e1}$ has the analogous form with the non-trivial 
entries (with positive signs in front of sinh's) in the 
$\{8th, 10th\}$ and (with negative signs in front of sinh's) in 
the $\{1st, 3rd\}$ columns and rows.  The third transformation 
${\bf\Omega}_{e2}$ has the non-trivial entries (with positive signs 
in front of sinh's) in the $\{3rd, 8th\}$ and (with negative signs 
in front of sinh's) in the $\{1st, 10th\}$ columns and rows.

The above three boosts $\delta_e, \delta_{e1}, \delta_{e2}$ 
induce the respective three $U(1)$ charges: $Q$ is the 
(electric) charge of the $U(1)$ gauge field strength dual to 
the field strength $H_{\mu\nu\rho}$ of the two-form field 
$B_{\mu\nu}$, while $Q_1^{(1)}$ and $Q_1^{(2)}$ are the 
(electric) charges of $A^{(1)}_{\mu\,1}$ gauge field (Kaluza-Klein 
gauge field, associated with the first compactified direction),
$A^{(2)}_{\mu\,1}$ gauge field (gauge field arising from the 
ten-dimensional two-form field, associated with the first 
compactified direction).
 
The final expression of the solution in terms of the (non-trivial)
five-dimensional bosonic fields is of the form
\footnote{The five-dimensional Newton's constant is taken to be 
$G_N^{D=5}={\pi\over 4}$ and we follow the convention of, {\it e.g.}, Ref. 
\cite{MP} for the definitions of the ADM mass, charges and 
angular momenta.}:  
\begin{eqnarray}
g_{11}&=&{{r^2+2m{\rm sinh}^2 \delta_{e1}+l^2_1{\rm cos}^2\theta+
l^2_2{\rm sin}^2\theta}\over{r^2+2m{\rm sinh}^2 \delta_{e2}+l^2_1
{\rm cos}^2\theta+l^2_2{\rm sin}^2\theta}}, \cr
e^{2\varphi}&=&{{(r^2+2m{\rm sinh}^2 \delta_{e}+l^2_1{\rm cos}^2
\theta+l^2_2{\rm sin}^2\theta)^2}\over{(r^2+2m{\rm sinh}^2
\delta_{e2}+l^2_1{\rm cos}^2\theta+l^2_2{\rm sin}^2\theta)
(r^2+2m{\rm sinh}^2 \delta_{e1}+l^2_1{\rm cos}^2\theta+l^2_2
{\rm sin}^2\theta)}},  \cr
A^{(1)}_{t\,1}&=&{{m{\rm cosh}\delta_{e1}{\rm sinh}\delta_{e1}}
\over{r^2+2m{\rm sinh}^2\delta_{e1}+l^2_1{\rm cos}^2\theta +
l^2_2{\rm sin}^2 \theta}}, \cr
A^{(1)}_{\phi\,1}&=&m{\rm sin}^2\theta
{{l_1{\rm sinh}\delta_{e1}{\rm sinh}\delta_{e2}{\rm cosh}\delta_{e}
-l_2{\rm cosh}\delta_{e1}{\rm cosh}\delta_{e2}{\rm sinh}\delta_{e}}
\over{r^2+2m{\rm sinh}^2\delta_{e1}+l^2_1{\rm cos}^2\theta +
l^2_2{\rm sin}^2 \theta}}, \cr
A^{(1)}_{\psi\,1}&=&m{\rm cos}^2\theta
{{l_1{\rm cosh}\delta_{e1}{\rm sinh}\delta_{e2}{\rm sinh}\delta_{e}
-l_2{\rm sinh}\delta_{e1}{\rm cosh}\delta_{e2}{\rm cosh}\delta_{e}}
\over{r^2+2m{\rm sinh}^2\delta_{e1}+l^2_1{\rm cos}^2\theta +
l^2_2{\rm sin}^2 \theta}}, \cr
A^{(2)}_{t\,1}&=&{{m{\rm cosh}\delta_{e2}{\rm sinh}\delta_{e2}}
\over{r^2+2m{\rm sinh}^2\delta_{e2}+l^2_1{\rm cos}^2\theta +
l^2_2{\rm sin}^2 \theta}}, \cr
A^{(2)}_{\phi\,1}&=&m{\rm sin}^2\theta
{{l_1{\rm cosh}\delta_{e1}{\rm sinh}\delta_{e2}{\rm cosh}\delta_{e}
-l_2{\rm sinh}\delta_{e1}{\rm cosh}\delta_{e2}{\rm sinh}\delta_{e}}
\over{r^2+2m{\rm sinh}^2\delta_{e2}+l^2_1{\rm cos}^2\theta +
l^2_2{\rm sin}^2 \theta}}, \cr
A^{(2)}_{\psi\,1}&=&m{\rm cos}^2\theta
{{l_1{\rm sinh}\delta_{e1}{\rm cosh}\delta_{e2}{\rm sinh}\delta_{e}
-l_2{\rm cosh}\delta_{e1}{\rm sinh}\delta_{e2}{\rm cosh}\delta_{e}}
\over{r^2+2m{\rm sinh}^2\delta_{e2}+l^2_1{\rm cos}^2\theta +
l^2_2{\rm sin}^2 \theta}}, \cr
B_{t\phi}&=&-2m\sin^2\theta(l_1\sinh\delta_{e1}\sinh\delta_{e2}\cosh
\delta_e-l_2\cosh\delta_{e1}\cosh\delta_{e2}\sinh\delta_e)(r^2+l^2_1
\cos^2\theta \cr
& &\ \ +l^2_2\sin^2\theta+m\sinh^2\delta_{e1}+m\sinh^2\delta_{e2})/ 
[(r^2+l^2_1\cos^2\theta+l^2_2\sin^2\theta+2m\sinh^2\delta_{e1}) \cr 
& & \ \ \times(r^2+l^2_1\cos^2\theta+l^2_2\sin^2\theta +
2m\sinh^2\delta_{e2})], \cr
B_{t\psi}&=&-2m\cos^2\theta(l_2\sinh\delta_{e1}\sinh\delta_{e2}
\cosh\delta_e-l_1\cosh\delta_{e1}\cosh\delta_{e2}\sinh\delta_e)
(r^2+l^2_1\cos^2\theta \cr
& & \ \ +l^2_2\sin^2\theta+m\sinh^2\delta_{e1}+m\sinh^2\delta_{e2})/ 
[(r^2+l^2_1\cos^2\theta+l^2_2\sin^2\theta+2m\sinh^2\delta_{e1}) \cr 
& &\ \ \times(r^2+l^2_1\cos^2\theta+l^2_2\sin^2\theta+
2m\sinh^2\delta_{e2})], \cr
B_{\phi\psi}&=&{{2m\cosh\delta_e\sinh\delta_e\cos^2\theta\sin^2\theta
(r^2+l^2_1\cos^2\theta+l^2_2\sin^2\theta+m\sinh^2\delta_{e1}+
m\sinh^2\delta_{e2})} \over {(r^2+l^2_1\cos^2\theta+l^2_2\sin^2\theta+
2m\sinh^2\delta_{e1})(r^2+l^2_1\cos^2\theta+l^2_2\sin^2\theta+
2m\sinh^2\delta_{e2})}}, \cr
ds^2_E&=& \bar{\Delta}^{1\over 3} \left [-{{(r^2+l^2_1{\rm cos}^2
\theta +l^2_2{\rm sin}^2\theta)(r^2+l^2_1{\rm cos}^2\theta 
+l^2_2{\rm sin}^2\theta-2m)}\over \bar{\Delta}} dt^2 \right .
\cr
&+&{r^2 \over {(r^2+l^2_1)(r^2+l^2_2)-2mr^2}} dr^2 +d\theta^2 
+{{4m{\rm cos}^2\theta {\rm sin}^2\theta}\over \bar{\Delta}}
[l_1 l_2\{(r^2+l^2_1{\rm cos}^2\theta +l^2_2{\rm sin}^2\theta) 
\cr
&-&2m({\rm sinh}^2\delta_{e1}{\rm sinh}^2\delta_{e2}+
{\rm sinh}^2\delta_{e}{\rm sinh}^2\delta_{e1}
+{\rm sinh}^2\delta_{e}{\rm sinh}^2\delta_{e2})\}+2m\{(l^2_1+l^2_2)
\cr
&\times&{\rm cosh}\delta_{e1}{\rm cosh}
\delta_{e2}{\rm cosh}\delta_{e}
{\rm sinh}\delta_{e1}{\rm sinh}\delta_{e2}{\rm sinh}\delta_{e}
-2l_1l_2{\rm sinh}^2 \delta_{e1}{\rm sinh}^2 \delta_{e2}
{\rm sinh}^2 \delta_{e}\}] d\phi d\psi 
\cr
&-&{{4m{\rm sin}^2\theta}\over \bar{\Delta}}[(r^2+l^2_1
{\rm cos}^2\theta+l^2_2{\rm sin}^2\theta)(l_1{\rm cosh}
\delta_{e1}{\rm cosh}\delta_{e2}{\rm cosh}\delta_{e}-
l_2{\rm sinh}\delta_{e1}{\rm sinh}\delta_{e2}{\rm sinh}\delta_{e})
\cr
&+&2ml_2{\rm sinh}\delta_{e1}{\rm sinh}\delta_{e2}{\rm sinh}
\delta_{e}]d\phi dt -{{4m{\rm cos}^2\theta}\over\bar{\Delta}}
[(r^2+l^2_1{\rm cos}^2\theta+l^2_2{\rm sin}^2\theta)
\cr
&\times&(l_2{\rm cosh}\delta_{e1}{\rm cosh}\delta_{e2}
{\rm cosh}\delta_{e}-l_1{\rm sinh}\delta_{e1}{\rm sinh}
\delta_{e2}{\rm sinh}\delta_{e})+2ml_1{\rm sinh}\delta_{e1}
{\rm sinh}\delta_{e2}{\rm sinh}\delta_{e}] d\psi dt 
\cr 
&+&{{{\rm sin}^2\theta}\over \bar{\Delta}} 
[(r^2+2m{\rm sinh}^2\delta_e+l^2_1)(r^2+2m{\rm sinh}^2\delta_{e1} 
+l^2_1{\rm cos}^2\theta+l^2_2{\rm sin}^2\theta)(r^2+2m{\rm sinh}^2
\delta_{e2} \cr
&+&l^2_1{\rm cos}^2\theta+l^2_2
{\rm sin}^2\theta)+2m{\rm sin}^2 \theta\{(l^2_1{\rm cosh}^2\delta_m 
-l^2_2{\rm sinh}^2\delta_m)(r^2+l^2_1{\rm cos}^2\theta +l^2_2{\rm sin}^2
\theta)\cr
&+&4ml_1 l_2{\rm cosh}\delta_{e1}
{\rm cosh}\delta_{e2}{\rm cosh}\delta_{e}{\rm sinh}\delta_{e1}
{\rm sinh}\delta_{e2}{\rm sinh}\delta_{e}-2m{\rm sinh}^2\delta_{e1}
{\rm sinh}^2\delta_{e2} \cr
&\times&(l^2_1{\rm cosh}^2\delta_e+l^2_2{\rm sinh}^2\delta_e)
-2ml^2_2{\rm sinh}^2\delta_e({\rm sinh}^2\delta_{e1}+
{\rm sinh}^2\delta_{e2})\}]d\phi^2 
\cr
&+&{{{\rm cos}^2\theta}\over \bar{\Delta}}[(r^2+2m{\rm sinh}^2
\delta_e+l^2_2)(r^2+2m{\rm sinh}^2\delta_{e1}+l^2_1{\rm cos}^2\theta
+l^2_2{\rm sin}^2\theta)(r^2+2m{\rm sinh}^2\delta_{e2}
\cr
&+&l^2_1{\rm cos}^2\theta+l^2_2{\rm sin}^2\theta)
+2m{\rm cos}^2 \theta\{(l^2_2{\rm cosh}^2\delta_e 
-l^2_1{\rm sinh}^2\delta_e)(r^2+l^2_1{\rm cos}^2\theta 
+l^2_2{\rm sin}^2\theta)
\cr
&+&4ml_1 l_2 {\rm cosh}\delta_{e1}
{\rm cosh}\delta_{e2}{\rm cosh}\delta_{e}{\rm sinh}\delta_{e1}
{\rm sinh}\delta_{e2}{\rm sinh}\delta_{e}-2m 
{\rm sinh}^2\delta_{e1}{\rm sinh}^2\delta_{e2}
\cr
&\times&\left .(l^2_1{\rm sinh}^2\delta_e+l^2_2{\rm cosh}^2\delta_e)
-2ml^2_1{\rm sinh}^2\delta_e({\rm sinh}^2\delta_{e1}+{\rm sinh}^2
\delta_{e2})\}]d\psi^2 \right], 
\label{5daxisol}
\end{eqnarray}
where
\begin{eqnarray}
\bar{\Delta} &\equiv& (r^2+2m{\rm sinh}^2\delta_{e1}+
l^2_1{\rm cos}^2\theta+l^2_2{\rm sin}^2\theta)(r^2+2m
{\rm sinh}^2\delta_{e2}+l^2_1{\rm cos}^2\theta+l^2_2
{\rm sin}^2\theta)\cr
&\times&(r^2+2m{\rm sinh}^2\delta_{e}+l^2_1{\rm cos}^2\theta
+l^2_2{\rm sin}^2\theta), 
\label{5def}
\end{eqnarray}
and the subscript $E$ in the line element denotes the 
Einstein-frame.  (In Eq. (\ref{5dkerr}), it was not necessary to 
distinguish between the string- and the Einstein-frames since 
in the case of the Kerr solution the dilaton is zero.)  The above 
solutions are obtained from the corresponding three-dimensional 
boosted solutions by first imposing the duality transformation 
(\ref{dual}) and then by constructing the five-dimensional fields, 
using the field redefinitions in section IIA. 
The $U(1)$ charges $Q's$, the ADM mass $M$  and the angular momenta  
$J's$ are given by:
\begin{eqnarray}
Q^{(1)}_1&=&2m{\rm cosh}\delta_{e1}{\rm sinh}\delta_{e1}, \ \ 
Q^{(2)}_1=2m{\rm cosh}\delta_{e2}{\rm sinh}\delta_{e2}, \ \ 
Q=2m{\rm cosh}\delta_{e}{\rm sinh}\delta_{e}, \cr
M&=&2m({\rm cosh}^2\delta_{e1}+{\rm cosh}^2\delta_{e2}+
{\rm cosh}^2\delta_{e})-3m\cr
&=&\sqrt{m^2+(Q^{(1)}_1)^2}+\sqrt{m^2+(Q^{(2)}_1)^2}+
\sqrt{m^2+Q^2}, \cr
J_{\phi}&=&4m(l_1{\rm cosh}\delta_{e1}{\rm cosh}\delta_{e2}
{\rm cosh}\delta_{e}-l_2{\rm sinh}\delta_{e1}{\rm sinh}\delta_{e2}
{\rm sinh}\delta_{e}), \cr
J_{\psi}&=&4m(l_2{\rm cosh}\delta_{e1}{\rm cosh}\delta_{e2}
{\rm cosh}\delta_{e}-l_1{\rm sinh}\delta_{e1}{\rm sinh}\delta_{e2}
{\rm sinh}\delta_{e}).
\label{5ddef}
\end{eqnarray}
Note that solution has the inner and outer horizons at:
\begin{equation}
r_{\pm}^2=m-{1\over 2}l_1^2-{1\over 2}l_2^2 \pm
{1\over 2}\sqrt{(l^2_1-l^2_2)^2+4m(m-l^2_1-l^2_2)},
\end{equation}
provided $m\ge (|l_1|+|l_2|)^2$.

The generating solution effectively corresponds to a 
six-dimensional target space background.  It would be interesting 
to map this solution onto a $\sigma$-model action with the 
corresponding six-dimensional target space specified by the 
background fields (\ref{5daxisol}).

The above solution with $Q=Q_{1}^{(1)}=Q_1^{(2)}$, {\it i.e.},
$\delta_e=\delta_{e1}=\delta_{e2}$, corresponds to the special case  
where both the dilaton $\varphi$ and the internal-circle-modulus 
field $g_{11}$ are constant. The solution with $Q=Q_1^{(2)}$, {\it i.e.},
$\delta_e=\delta_{e2}$, corresponds to the case  
where the six-dimensional dilaton $\varphi_6=\varphi+{\textstyle{1\over
2}}\log {\rm det} g_{11}$ is constant. In this case, 
 with a  subsequent rescaling of the 
asymptotic values of the  scalar fields
 one obtains  the static solutions studied in Ref. \cite{HS} and rotating 
solutions studied in Ref.\cite{BLMPSV}. 

\subsection{$T$-Duality Transformations}

The subset of the $[SO(5)\times SO(21)]/[SO(4)\times SO(20)]$ 
transformations yields the following charge assignments: 
\begin{equation}
\vec{\cal Q}^{\prime} = {1\over \sqrt{2}} {\cal U}^T \left ( 
\matrix{U_5 ({\bf e}_u-{\bf e}_d)\cr U_{21}\left(\matrix{{\bf e}_u
+{\bf e}_d \cr 0_{16}}\right)}\right ), 
\label{newch}
\end{equation}
where 
\begin{equation}
{\bf e}^T_u \equiv (Q^{(1)}_1 ,\overbrace{0,...,0}^{4}),\ \ \ 
{\bf e}^T_d \equiv (Q^{(2)}_1, \overbrace{0,...,0}^{4}),
\label{submat}
\end{equation}
where $U_5 \in SO(5)/SO(4)$, $U_{21} \in SO(21)/SO(20)$, 
$0_{16}$ is a $(16\times 1)$-matrix with zero entries, and 
${\cal U} \equiv \left ( \matrix{{1\over\sqrt{2}}I_5&
-{1\over\sqrt{2}}I_5&0\cr {1\over\sqrt{2}}I_5&{1\over\sqrt{2}}I_5&
0\cr 0&0&I_{16}} \right )$.  The moduli field  matrix $M$ is 
transformed to
\begin{equation}
M^{\prime}={\cal U}^T \left(\matrix{U_5 &0 \cr 0 & U_{21}}\right)
{\cal U}M{\cal U}^T \left (\matrix{U^T_5 & 0 \cr 0 & U^T_{21}}
\right){\cal U},
\label{genscalar}
\end{equation}
The subsequent transformation by a constant $O(5,21)\times SO(1,1)$ 
matrix allows one to write the solution in terms of the arbitrary 
asymptotic values $M_\infty$ and $\varphi_\infty$.

\section{BPS-Saturated Limit and Deviations from it}

\subsection{BPS-Saturated Limit}

The BPS-limit (of the generating solution) is obtained by  
taking $m\to 0$, while keeping the three charges
$Q_1^{(1)}$, $Q^{(2)}_1$ and $Q$, as well as $J_\phi$ and 
$J_\psi$ finite.  This is 
achieved by taking the following 
limits: $m\to 0$, $l_{1,2}\to 0$ and $\delta_{e1,e2,e}\to\infty$ while 
keeping ${1\over 2}me^{2\delta_{e1}}=Q^{(1)}_1,\,{1\over 2}
me^{2\delta_{e2}}=Q^{(2)}_1,\,{1\over 2}me^{2\delta_{e}}=Q$,\,
$l_1/m^{1/2}=L_1$, and $l_2/m^{1/2}=L_2$ constant
\footnote{Since we choose the boost parameters to be positive, 
{\it i.e.}, $\delta\to +\infty$, the $U(1)$ charges are positive.  
Had we taken the boost parameters to be negative, the $U(1)$ charges 
are negative but they will appear with absolute values in the 
solutions.}. 
The BPS-saturated solution is then of the following form:  
\begin{eqnarray}
A^{(1)}_{t\,1}&=&{{{1\over 2}Q^{(1)}_1} \over {r^2+Q^{(1)}_1}}, 
\ \ \ \ 
A^{(1)}_{\phi\,1}={{{1\over 4}J\sin^2\theta} \over 
{r^2+Q^{(1)}_1}}, \ \ \ \ 
A^{(1)}_{\psi\,1}={{{1\over 4}J\cos^2\theta} \over 
{r^2+Q^{(1)}_1}}, \cr
A^{(2)}_{t\,1}&=&{{{1\over 2}Q^{(2)}_1} \over {r^2+Q^{(2)}_1}}, 
\ \ \ \ 
A^{(2)}_{\phi\,1}={{{1\over 4}J\sin^2\theta} \over 
{r^2+Q^{(2)}_1}}, \ \ \ \ 
A^{(2)}_{\psi\,1}={{{1\over 4}J\cos^2\theta} \over 
{r^2+Q^{(2)}_1}}, \cr
B_{t\phi}&=&-{{{1\over 2}J\sin^2\theta(r^2+{1\over 2}Q^{(1)}_1+
{1\over 2}Q^{(2)}_1)}\over {(r^2+Q^{(1)}_1)
(r^2+Q^{(2)}_1)}}, \ \ \ \ \ 
B_{t\psi}={{{1\over 2}J\cos^2\theta(r^2+{1\over 2}Q^{(1)}_1+
{1\over 2}Q^{(2)}_1)}\over {(r^2+Q^{(1)}_1)
(r^2+Q^{(2)}_1)}}, \cr
B_{\phi\psi}&=&{{Q\cos^2\theta\sin^2\theta 
(r^2+{1\over 2}Q^{(1)}_1+{1\over 2}Q^{(2)}_1)} \over 
{(r^2+Q^{(1)}_1)(r^2+Q^{(2)}_1)}}, \ \ 
g_{11}={{r^2+Q^{(1)}_1}\over{r^2+Q^{(2)}_1}}, 
\ \  e^{\varphi}={{(r^2+Q)}\over{[(r^2+Q^{(1)}_1)
(r^2+Q^{(2)}_1)]^{1\over 2}}}, \cr
ds^2_E&=&\bar{\Delta}^{1\over 3}\left[-{r^4\over \bar{\Delta}}dt^2 
+{{dr^2} \over r^2} +d\theta^2+{{J^2{\rm cos}^2\theta
{\rm sin}^2\theta}\over {2\bar{\Delta}}}d\phi d\psi-
{{2Jr^2{\rm sin}^2\theta}\over 
\bar{\Delta}}dtd\phi+{{2Jr^2{\rm cos}^2\theta}\over\bar{\Delta}}
dtd\psi \right.\cr
&+&{{{\rm sin}^2\theta} \over \bar{\Delta}}\{(r^2+Q^{(1)}_1)
(r^2+Q^{(2)}_1)(r^2+Q)-{1\over 4}J^2{\rm sin}^2\theta\}
d\phi^2 \cr
&+&\left.{{{\rm cos}^2\theta} \over \bar{\Delta}}\{(r^2+Q^{(1)}_1)
(r^2+Q^{(2)}_1)(r^2+Q)-{1\over 4}J^2{\rm cos}^2
\theta\}d\psi^2 \right], 
\label{5dBPS}
\end{eqnarray}
where
\begin{equation}
\bar{\Delta}\equiv (r^2+Q^{(1)}_1)(r^2+Q^{(2)}_1)(r^2+Q).\ \ \ 
\label{5dBPSdef}
\end{equation}
The solution is specified by three charges and {\it only one} 
rotational parameter $J$
\footnote{In the case one takes one or three boost parameters 
to be negative, one obtains the BPS limit with $J_{\phi}=J_{\psi}$.}:
\begin{equation}
J_\phi=-J_\psi\equiv J=(2Q^{(1)}_1Q^{(2)}_1Q)^{1\over 2}
(L_1-L_2),
\end{equation}
while its ADM mass saturates the Bogomol'nyi bound:
\begin{equation}
M_{BPS}=Q^{(1)}_1+Q^{(2)}_2+Q.
\end{equation}
The surface area of the horizon is of the form:
\begin{equation}
A_{BPS}=4\pi^2[(Q^{(1)}_1Q^{(2)}_1Q)(1-{1\over 2}(L_1-L_2)^2)]^{1\over 2}
=4\pi^2[Q^{(1)}_1Q^{(2)}_1Q-{1\over 4}J^2]^{1\over 2}.
\label{5dbpsarea}
\end{equation}
The above solution with $Q=Q_{1}^{(1)}=Q_1^{(2)}$ and the 
subsequently rescaled asymptotic value of the dilaton 
$\varphi_\infty\ne 0$ was obtained for the static BPS-saturated 
states in Ref. \cite{SV} and for rotating ones in Ref. \cite{BMPV}. 
The above general form of the solution (with all three charges
different) was given in Ref. \cite{T} as a solution 
of the corresponding conformal $\sigma$-model
\footnote{Related $\sigma$-models, describing more general 
5-dimensional rotating solutions, are studied in Ref. 
\cite{TII}.}.   

The mass of the BPS-saturated solution (at generic points of the 
moduli and the coupling space) can be obtained by acting with
the subset of $[SO(5)\times SO(21)]/[SO(4)\times SO(20)]$ 
transformations and then undoing the asymptotic values of moduli 
$M_{\infty}=I_{26}$ and the dilaton $\varphi_\infty=0$ with the 
constant $O(5,21)\times SO(1,1)$ transformations, as spelled out 
in Section  III A.  The procedure yields the following form of 
the ADM mass
\begin{equation}
M_{BPS}= e^{2\varphi_{\infty}/3}
[{\vec \alpha}^T({\cal M}_{\infty}+{L}){\vec \alpha}]^{1/2} +
e^{-4\varphi_{\infty}/3} \beta,  
\label{genmass}\end{equation}
while the macroscopic entropy  ${\bf S}=A/(4G_N^{D=5})$
(here $A$ is the area of the horizon at $r_+=r_-=0$) is of the form:
\begin{equation}
{\bf S}_{BPS}=4\pi\sqrt{\beta({\vec\alpha}^T{ L}{\vec \alpha})-
{1\over 4}J^2},
\label{genarea}
\end{equation}
where the expression has been written in terms of conserved, 
quantized charges $\beta$ and ${\vec \alpha}$, which are related  
to the physical charges $Q$ and $\vec {\cal Q}$ (\ref{newch}) 
in the following way:
\begin{equation}
Q=e^{-4\varphi_\infty/3}\beta, \ \ \ \ \   
\vec{\cal  Q}=e^{2\varphi_\infty/3}M_\infty \vec{\alpha}. 
\end{equation} 
While the mass depends on the asymptotic values of 
moduli and the dilaton coupling, the entropy for the BPS states 
is a universal quantity \cite{LW,CT,S,FK}, and thus depends only 
on conserved (quantized) charges $\beta$ and $\vec \alpha$.  

\subsection{Deviations from the BPS-Saturated Limit}

It is useful to obtain the explicit form of the solution that is 
infinitesimally close to the BPS limit, since in certain specific 
cases the microscopic entropy can be calculated for such solutions 
as well \cite{HS,BLMPSV}.

The deviation from the BPS limit of the generating solution
is achieved by taking $m$ small, but non-zero, while keeping 
other parameters $Q_1^{(1)}, Q_1^{(2)}, Q, L_1$, and $L_2$ finite.

We have pursued the expansion of the solution up to the 
linear order in $m$.  In particular, now, the solution has the inner 
and outer horizons at:
\begin{equation}
r^2_{\pm}=m\left(1-{1\over 2}(L^2_1+L^2_2)\pm{1\over 2}
\sqrt{[2-(L_1+L_2)^2]
[2-(L_1-L_2)^2]}\right).
\label{devhorizon}
\end{equation}
The area of the  outer horizon $r_+$ can be written as:
\begin{eqnarray}
A&=&4\pi^2\left [(Q^{(1)}_1Q^{(2)}_1Q)\sqrt{1-{1\over 2}(L_1-L_2)^2}
+m(Q^{(1)}_1Q^{(2)}_1+Q^{(1)}_1Q+Q^{(2)}_1Q)\right.\cr
&\times&\left.\sqrt{[1-{1\over 2}(L_1+L_2)^2]}\right]^{1\over 2}, 
\label{devarea}
\end{eqnarray}
which is correct in the expansion up to the leading order in $m$, 
only.  The rotational parameters $J_\phi$ and $J_\psi$ are not 
equal in magnitude and opposite in sign anymore, but are of the form:
\begin{eqnarray}
J&\equiv&\textstyle{1\over 2}(J_\phi-J_\psi)= 
(2Q^{(1)}_1Q^{(2)}_1Q)^{1\over 2}(L_1-L_2)+{\cal O}(m^2),\cr
\Delta J&\equiv&{\textstyle{1\over 2}}(J_\psi+J_\phi)= 
m(2Q^{(1)}_1Q^{(2)}_1Q)^{1\over 2}\left(
{1\over{Q^{(1)}_1}}+{1\over{Q^{(2)}_1}}+{1\over{Q}}\right)(L_1+L_2) 
+{\cal O}(m^2), \end{eqnarray}
while the ADM mass is still conveniently kept as:
\begin{equation}
M=\sqrt{m^2+(Q^{(1)}_1)^2}+\sqrt{m^2+(Q^{(2)}_1)^2}+
\sqrt{m^2+Q^2}.
\end{equation}
Note that in the case where one of the charges is taken small, 
{\it e.g.}, $Q^{(1)}_1\to 0$, as in the recent study of the 
microscopic entropy near the BPS-saturated limit \cite{HS,BLMPSV},
the ADM mass is $M=M_{BPS} +{\cal O}(m)$, while the area is 
$A={\cal O}(m^{1/2})$.  However, when all the charges are taken 
non-zero the deviation from the BPS limit is of the form 
$M=M_{BPS} +{\cal O}(m^2)$ and $A=A_{BPS}+ {\cal O}(m)$.  
In particular, further study of the microscopic entropy for
an infinitesimal deviation from the general BPS-saturated limit 
is needed.

\acknowledgments
We would like to thank C. Hull, F. Larsen and especially  
A.A. Tseytlin for useful discussions.  The work is supported 
by the Institute for Advanced Study funds and J. Seward Johnson 
foundation, U.S. DOE Grant No. DOE-EY-76-02-3071, the NATO 
collaborative research grant CGR No. 940870 and the National 
Science Foundation Career Advancement Award No. PHY95-12732.

\end{document}